# Investigations on Important Properties of the 10 cm × 10 cm GEM Prototype

*Kiadtisak Saenboonruang\*, Piyakul Kumphiranon,*

*Kittipong Kulasri and Anawat Ritthirong*

*Department of Applied Radiation and Isotopes, Faculty of Science, Kasetsart University, Bangkok, 10900, Thailand*

The Gas Electron Multiplier (GEM) detector is one of promising particle and radiation detectors that has been improved greatly from previous gas detectors. The improvement includes better spatial resolutions, higher detection rate capabilities, and flexibilities in designs. In particular, the 10 cm × 10 cm GEM prototype is designed and provided by the Gas Detectors Development group (GDD) at CERN, Switzerland. With its simplicity in operations and designs, while still maintaining high qualities, the GEM prototype is suitable for both start-up and advanced researches. This article aims to report the investigations on some important properties of the 10 cm × 10 cm GEM detector using current measurement and signal counting. Results have shown that gains of the GEM prototype exponentially increase as voltage supplied to the detector increases, while the detector reaches full efficiency (plateau region) when the voltage is greater than 4100 V. In terms of signal sharing between X and Y strips of the readout, X strips, which is on the top layer of the readout, collect ~57% of the total signal. For the uniformity test, the GEM prototype has slightly higher efficiencies at the center of the detector and decreases as positions are closer to edges. These results can be used for future references and for better understanding in the GEM prototype's characteristics.

Email: fscikssa@ku.ac.th

Fax: +66-2-579-5530

# I. INTRODUCTION

The Gas Electron Multiplier (GEM) detector was invented by F. Sauli in 1997 [1]. Since the invention, the GEM detector has gained attentions amongst international scientists and researchers. Parts of the success are from its improved properties from previous types of gas detectors. Examples of the improvements include [2]:

- the ability to operate in most gases
- the ability to vary gains of the detector (up to $10^5$)
- excellent spatial resolution (50 µm or better) [3]
- high rates capability ($10^5$ Hz/mm$^2$)
- flexibilities in designs
- relatively low cost.

The GEM detectors are now utilized in various scientific researches including tracking devices in particle and nuclear physics [4], medical imaging [5], astronomy [6], and neutron [7].

The 10 cm × 10 cm GEM prototype was designed, developed, and supplied by the Gas Detectors Development Group (GDD) at CERN. The triple GEM prototype consists of three GEM foils, which are 50-µm thin insulating foils made of polyimide (Kapton). Each foil is sandwiched by two thin copper plates. The GEM foil is perforated with arrays of 70-µm diameter holes (GEM holes) with 140-µm pitches between two adjacent holes. Voltage difference of 250-400 V is supplied to the two copper plates such that strong electric fields are formed inside the GEM holes. In addition to GEM foils, the drift cathode is usually made of a thin sheet of aluminized Kapton, where the aluminum side is supplied with the most negative voltage (~4000 V). All GEM foils and the drift cathode are enclosed in a gas-tight box with one gas inlet and one gas outlet. Readout of the GEM prototype has the XY configuration where two sets of 256 thin conducting wires running perpendicular to each other in X and Y directions. The schematic drawings of the GEM prototype and the readout strips are shown in Fig. 1 and Fig. 2. The widths of the X and Y strips are 50 µm and 150 µm respectively and the difference in strip widths is to improve signal sharing between X and Y strips. To operate the GEM detector, appropriate gas filling must flow throughout the detector. In principle, a pure noble gas such as argon can be used. However, in order to improve the stability of the detector, a gas mixture is usually used. The most common gas mixture is Ar/$CO_2$ with the ratio of 70:30. When ionizing particles and radiation pass through the GEM detector, they will ionize gas molecules inside the detector creating primary electron. These primary electrons will drift down to GEM foils and gain enough energy from strong electric fields inside GEM holes to further ionize gas molecules and number of electrons will be greatly amplified. The amplified signal will be captured by strips at the readout and will be transferred to appropriate DAQ system for data processing [8].

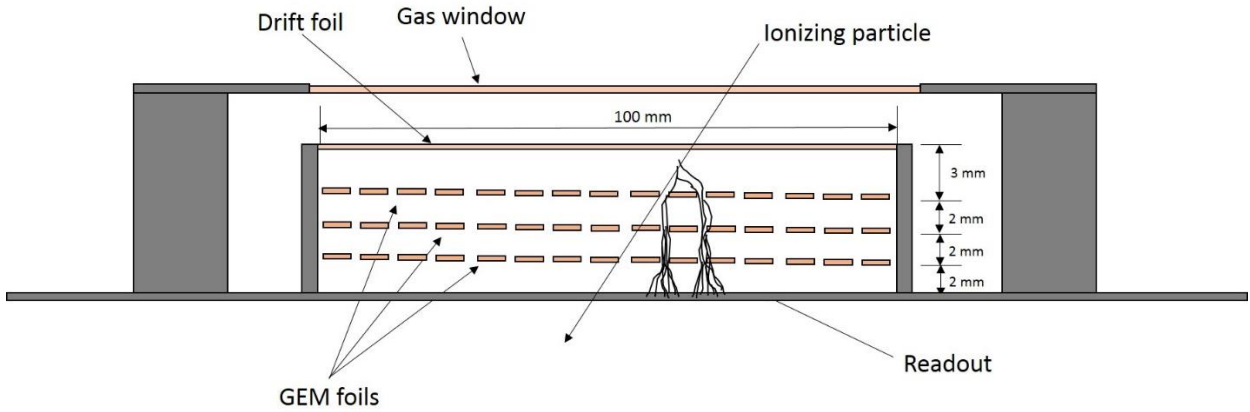

Fig. 1. The schematic drawing of the 10 cm × 10 cm GEM prototype. GEM foils and drift cathode are stacked at the center of the gas-tight box with the readout serving as the base of the detector

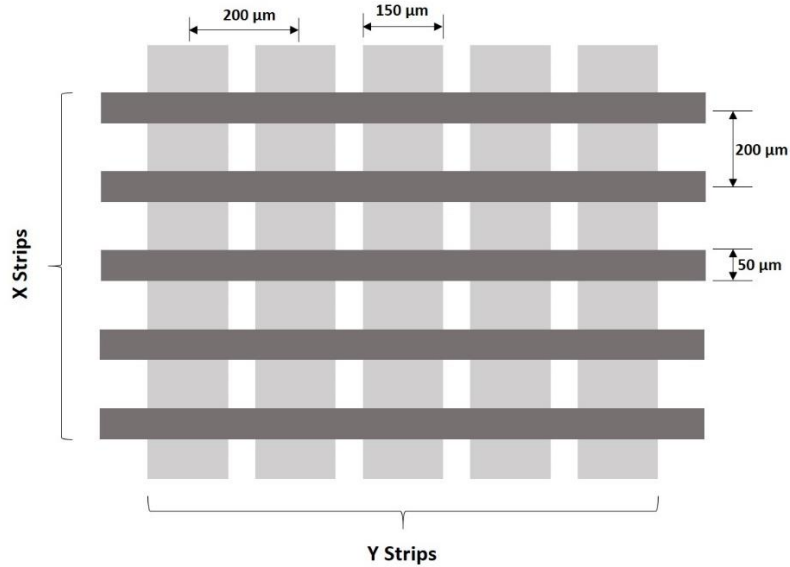

Fig. 2. The schematic drawing of the readout of the 10 cm × 10 cm GEM prototype in XY configuration. The X strips are wires that are placed on top with the strip width of 50 μm

and a distance between strips of 200 μm. The Y strips are wires that are placed at the bottom with the strip width of 150 μm and a distance between strips of 200 μm

The GEM technology has been developed greatly in recent years. Sophisticated designs and large-sized detectors have been manufactured to be used in many advanced researches. However, the GEM prototype still plays important roles in many researches, especially start-up researches and preliminary studies. Information and methods to investigate characteristics of the GEM prototype will be very useful for these researches, and hence, this article aims to report the investigations on main and important properties of the GEM prototype: the plateau behavior of the detector, gains of the detector as a function of the power supply voltages, signal sharing between X and Y strips, and the uniformity of the detector.

## II. MATERIALS AND METHODS

**Plateau investigation**

To investigate plateau region of the GEM prototype, rates of 5.9-keV X-ray from Fe-55 were measured as voltages of power supply varied from 3900 to 4300 V in 50-V increments. Setup schematic diagram of the rate measurement is shown in Fig. 3. The preamplifier used for this purpose was a charge sensitive amplifier (Cremat-110) with ×4 amplification [9]. The threshold at the discriminator was set at 65 mV to eliminate all electronic noises. The power was supplied to the GEM prototype through a voltage divider shown in Fig. 4.

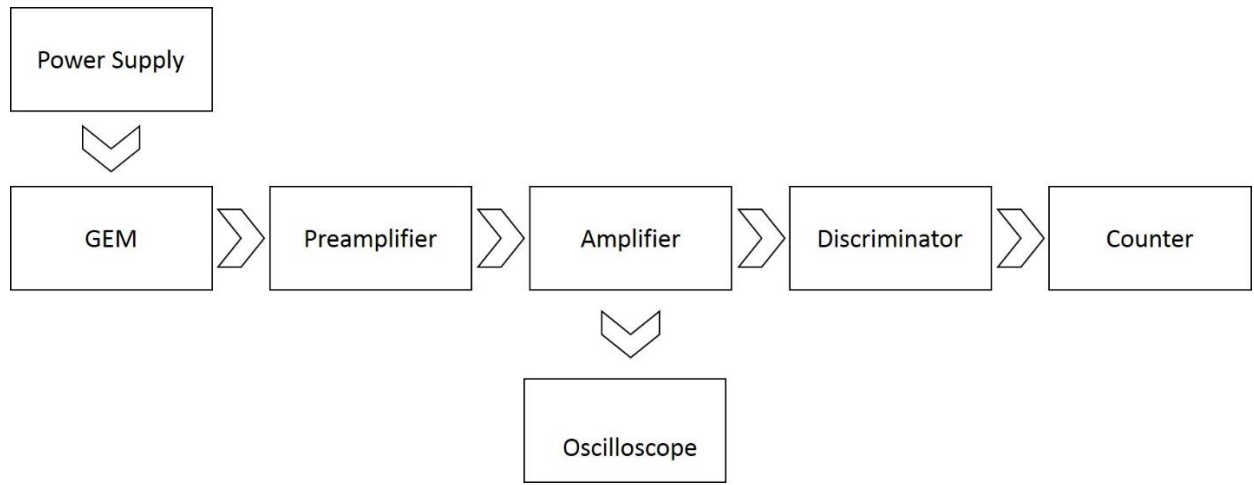

Fig. 3. The figure shows schematic drawing of the setup for the rate measurement in the plateau investigation

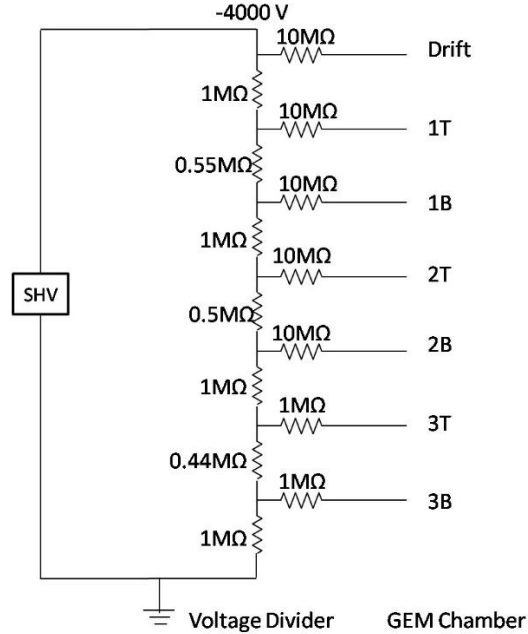

Fig. 4. The figure shows the schematic drawing of the voltage divider used for supplying voltages to the GEM prototype

**Gain measurement**

To measure gains of the GEM prototype, currents from 5.9-keV X-ray emitted from Fe-55 were measured as voltages of power supply varied from 3900 to 4300 V in 50-V increments. Consider the equation

$$I = R \times N \times G \times e \qquad (1)$$

where $I$ is the current, $R$ is the rate of the 5.9-keV X-ray from Fe-55, $N$ is the number of primary electrons, $G$ is the gain of the detector, and $e$ is the charge of an electron ($e = 1.6 \times 10^{-19}$ C). In order to obtain $G$, values of $I$, $R$, and $N$ must be carefully measured and evaluated. To measure $I$, a pico-ammeter which has a 20-fm current resolution was used for the current measurement. The setup for the current measurement is shown in Fig. 5. $R$ was the value of the rate of 5.9-keV X-ray emitted from Fe-55 at the plateau region. $N$ could be estimated using the average work function ($W$) of the gas mixture (Ar/$CO_2$) in the ratio of 70:30. The average $W$ was calculated using Eq. 2.

$$\frac{1}{W} = \frac{\%\ of\ Ar}{W_{Ar}} + \frac{\%\ of\ CO_2}{W_{CO_2}} \qquad (2)$$

where $W_{Ar}$ = 25 eV, $W_{CO_2}$ = 34 eV, *% of Ar* = 0.7, and *% of CO$_2$* = 0.3. Thus, $W$ = 27.8 eV [10]. Assuming that only photoelectric effect occurs during the interaction between X-ray and gas molecules, the value of *N* would be 212 electrons.

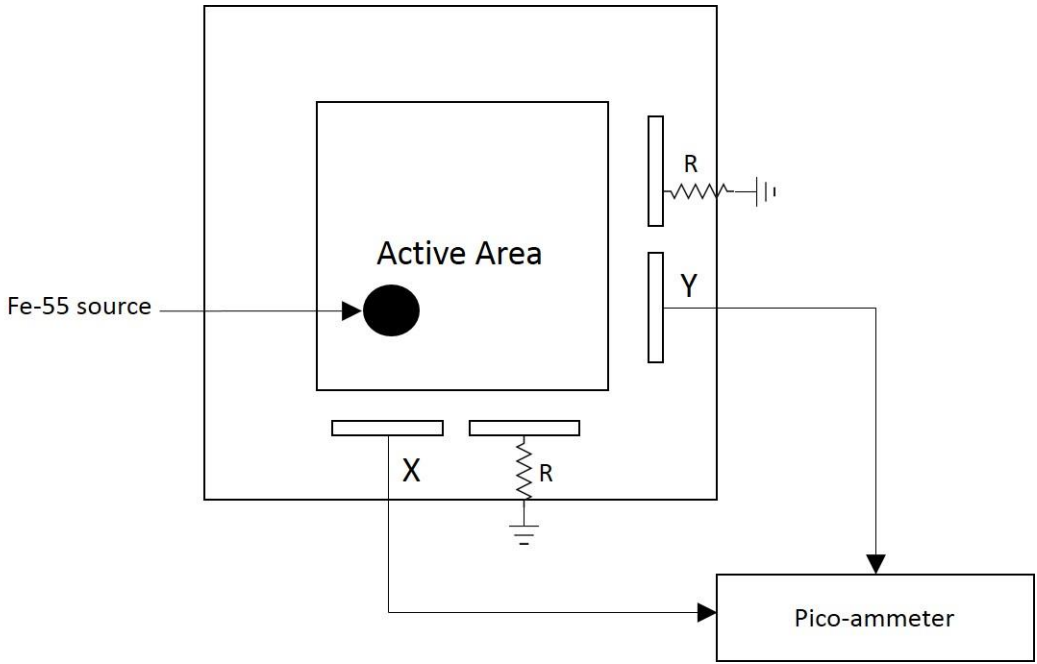

Fig. 5. The schematic drawing of the current measurement using pico-ammeter and a Fe-55 source. Two scenarios: only X strips and a combination of X and Y strips, were used for current measurement

**Signal sharing between X and Y strips**

Although the purpose of the different widths in the XY readout strips is to improve signal sharing between X and Y such that equal signals are shared between them, inequality in the signal sharing could still occur. To investigate the signal sharing, currents were measured in two scenarios; only X strips ($I_1$) and a combination of X and Y strips ($I_2$). The ratio of $I_1/I_2$ indicates percentages of the signal collected by X strips.

**Uniformity test**

Since efficiencies of the GEM detector at areas near edges of the active area are expected to be lower than efficiencies at the center of the active area, investigation of the uniformity of the GEM detector is needed to better understand these differences. To test the uniformity, the 10 cm

× 10 cm active area was divided into 36 positions (6 columns and 6 rows). Am-241, which emits primary alpha particles and 59-keV secondary gamma, was placed on each position. In order to correctly compare efficiencies from different positions, gas flow rate (3.0 L/hr), detection duration (3 minutes), and power supply voltage (4100 V) were set to be the same throughout the measurement. For each position, numbers of counts detected using the setup in Fig. 3 were collected and averaged. After completing all 36 positions, numbers of counts were plotted using the OriginPro software to produce a contour of uniformity.

## III. RESULTS AND DISCUSSION

**Plateau investigation**

Results of count rates as a function of power supply voltages are shown in Table 1 and Fig. 6.

Table 1. Count rates of the GEM prototype as the power supply voltages varied.

| Power Supply Voltage (V) | Count Rate (Hz) |
|:---:|:---:|
| 3900 | 270 |
| 3950 | 501 |
| 4000 | 611 |
| 4050 | 647 |
| 4100 | 665 |
| 4150 | 658 |
| 4200 | 665 |
| 4250 | 676 |
| 4300 | 673 |

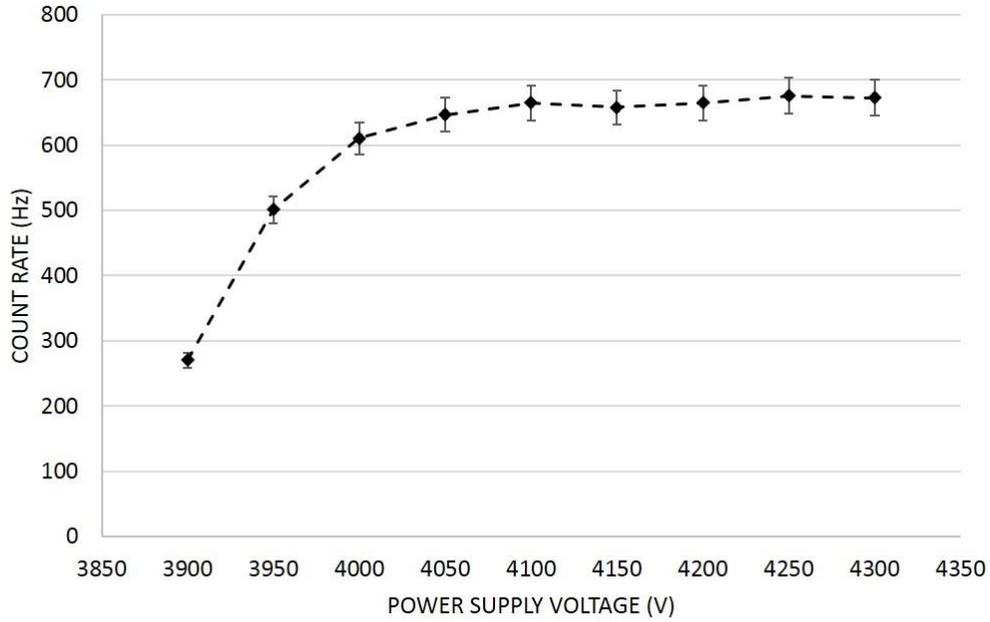

Fig. 6. Figure shows count rates of the GEM prototype as a function of power supply voltages. The GEM prototype reached plateau region after ~4100 V.

As shown in Fig. 6, the full efficiency (plateau region) of the GEM prototype occurred when the power supply voltage was higher than ~4100 V. This implies that even though amplitudes of signals became larger as the voltages increase, the full efficiency of the GEM prototype was already achieved at V = 4100 V.

**Gain measurement**

From previous calculations and results, $R \sim 670$ Hz, $N = 212$ electrons, and $e = 1.6 \times 10^{-19}$ C. The current values with a combination of X and Y strips as a function of power supply voltages are shown in Table 2.

Table 2. Currents of the GEM prototype as the power supply voltages varied

| Power Supply Voltage (V) | Current (nA) |
|---|---|
| 3900 | 0.05 |
| 3950 | 0.07 |
| 4000 | 0.10 |
| 4050 | 0.15 |
| 4100 | 0.20 |
| 4150 | 0.29 |
| 4200 | 0.42 |
| 4250 | 0.60 |
| 4300 | 0.90 |

With values of *I, N, e,* and *R* indicated above, gains for different voltages were calculated and are shown in Table 3 and Fig. 7.

Table 3. Gains of the GEM prototype as the power supply voltages varied

| Power Supply Voltage (V) | Gain |
|---|---|
| 3900 | 2200.08 |
| 3950 | 3080.12 |
| 4000 | 4400.17 |
| 4050 | 6600.25 |
| 4100 | 8800.34 |
| 4150 | 12760.49 |
| 4200 | 18480.71 |
| 4250 | 26401.01 |
| 4300 | 36901.52 |

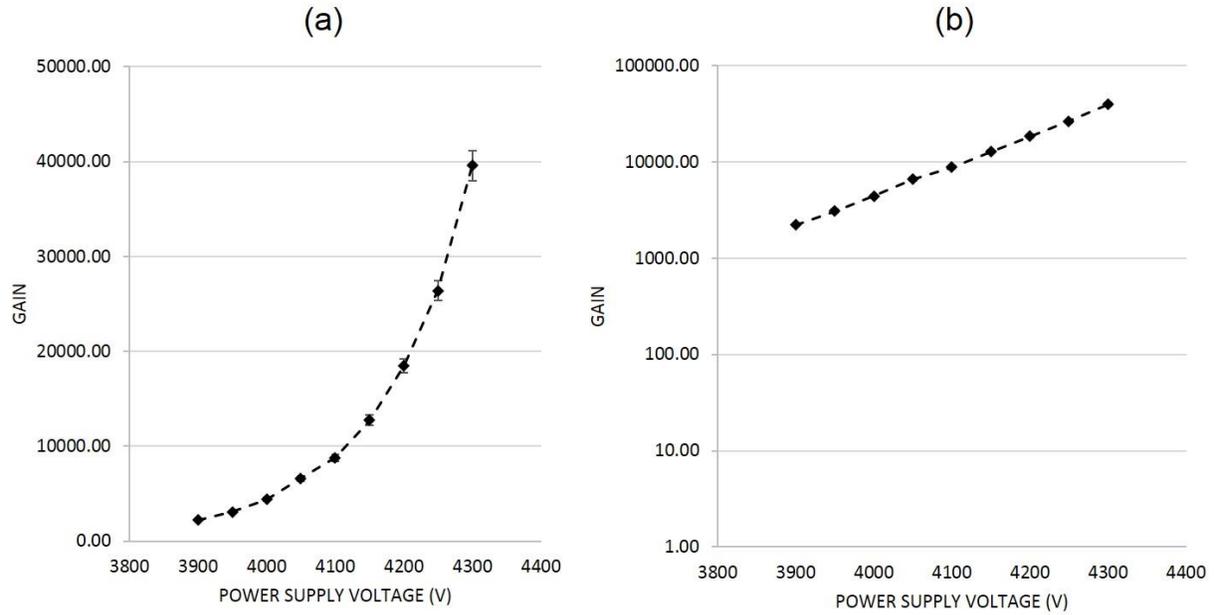

Fig. 7. Figures (a) and (b) show gains of the GEM prototype as power supply voltages varied. Figure (b) is plotted in logarithm scale.

As shown in Fig. 7(a) and Fig. 7(b), gains of the GEM prototype exponentially increase as the power supply voltages increase.

**Signal sharing between X and Y strips**

To determine ratio of signal sharing between X and Y strips, currents from X strips only ($I_1$) were compared with currents from a combination of X and Y strips ($I_2$). Values of currents from both scenarios are shown in Table 4. Values of $I_1/I_2$ are plotted in Fig. 8.

Table 4. Current measurement from X strips only and a combination of X and Y strips

| Power Supply Voltage (V) | Current from X Strips, $I_1$, (nA) | Current from A combination of X and Y strips, $I_2$, (nA) | $I_1/I_2$ |
|---|---|---|---|
| 3900 | 0.03 | 0.05 | 0.60 |
| 3950 | 0.04 | 0.07 | 0.57 |
| 4000 | 0.06 | 0.10 | 0.60 |
| 4050 | 0.08 | 0.15 | 0.53 |
| 4100 | 0.12 | 0.20 | 0.60 |
| 4150 | 0.16 | 0.29 | 0.55 |
| 4200 | 0.25 | 0.42 | 0.60 |
| 4250 | 0.33 | 0.60 | 0.55 |
| 4300 | 0.51 | 0.90 | 0.57 |

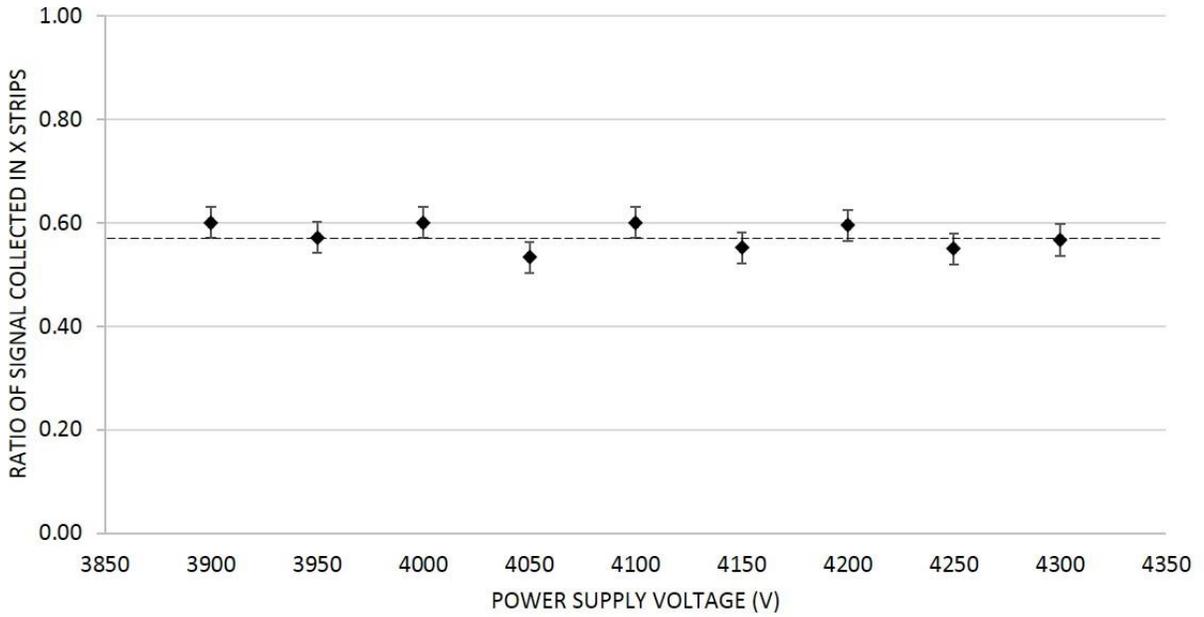

Fig. 8. The figure shows ratios of $I_1$ and $I_2$. The dotted line is the average of all ratios, which is 0.57±0.03.

Fig. 8 shows that the average $I_1/I_2 = 0.57\pm0.03$. Hence, the X trips, which is narrower and located on the top layer of the readout, collect larger signals compared to Y strips. To improve better signal sharing, a new design and better manufacture of the readout are required [4].

**Uniformity test**

Fig. 9 shows the uniformity of the GEM prototype using Am-241 as a gamma emitter. Areas near the center of the active area had higher efficiencies compared to areas near edges. This behavior is expected since ionizing particles or ionized electrons occurred near edges have possibilities to travel or drift out of the active area, and thus, lower its overall efficiencies and signal amplitudes. However, if considering areas with at least 1 cm away from edges, the efficiencies were well within 20% from each position.

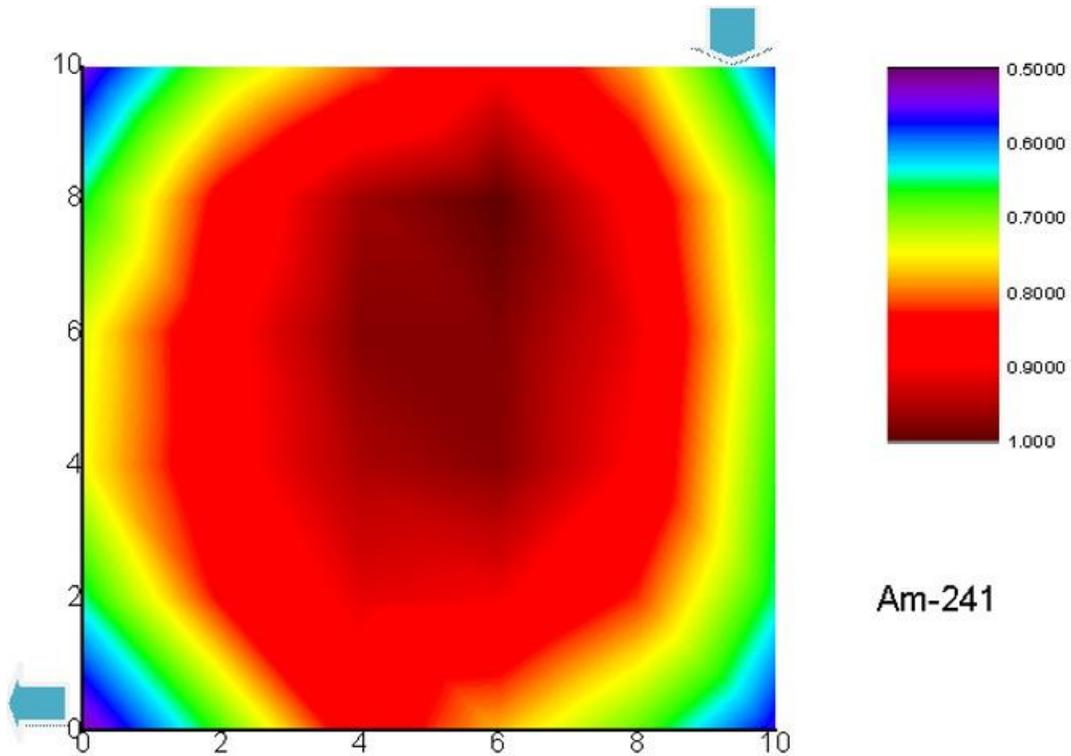

Fig. 9. The figure shows the uniformity of the GEM prototype. Higher efficiencies are clearly shown at the center of the detector

## IV. CONCLUSIONS

The GEM detector has become one of the most promising particle and radiation detectors nowadays. It has been utilized in various scientific researches including particle and nuclear physics, medical applications, astronomy, and national security. Since the 10 cm × 10 cm GEM prototype, which is designed by the GDD group at CERN, has simple designs and assemble procedures while still maintaining excellent properties, it is exceptionally suitable for both start-up and advanced researches. Many researchers have relied their researches to the excellences in properties of the GEM prototype. This article reports the investigations on main properties of the GEM prototype, which include

- Plateau: the GEM prototype reached full efficiency when the power supply voltages reached 4100 V and became relatively constant when the voltages were greater than 4100 V.
- Gain: gains of the GEM prototype increased exponentially with the increase in power supply voltages. The relationship between gains and voltages could be described by the equation $G = 2 \times 10^{-9} e^{0.0072V}$.
- Signal sharing: it was found that X strips, which are narrower and located on the top layer of the readout, collected larger signals than Y strips by ~30%.
- Uniformity: the GEM prototype had higher efficiencies at the center of the active area, while areas near edges had lower efficiencies.

These investigations are very useful for researchers to use as future references and to better understand behaviors of the GEM detector. Further researches on the GEM detector should be followed in order to improve and to widen possible applications.

## ACKNOWLEDGEMENT

We gratefully acknowledge a financial support from the Faculty of Science and Kasetsart University the grant no. RFG2-5.